\documentclass[aip,reprint,graphicx,twocolumn,amsmath,amssymb,floatfix]{revtex4-1}

\usepackage{graphicx}% Include figure files
\usepackage{appendix}
\usepackage[USenglish]{babel}
\usepackage[utf8]{inputenc}
\usepackage[T1]{fontenc}
\usepackage{mathptmx}
\usepackage{float}
\usepackage{epstopdf}
\usepackage{hyperref}

\usepackage{color}
\usepackage{tikz}
\usetikzlibrary{shapes}
\newcommand{\markerone}{\raisebox{-0.5pt}{\tikz{\node[draw,scale=0.8,circle,fill=blue!59!green,blue!59!green](){};}}}
\newcommand{\markerthree}{\raisebox{0.5pt}{\tikz{\node[draw,scale=0.5,regular polygon, regular polygon sides=3,fill=black!70!white,black!70!white,rotate=0](){};}}}
\newcommand{\markertwo}{\raisebox{-1pt}{\tikz{\node[draw,scale=0.6,diamond,aspect=0.75,fill=teal,teal](){};}}}
\newcommand{\Cross}{$\mathbin{\tikz [x=1.4ex,y=1.4ex,line width=.3ex, red] \draw (0,0) -- (1,1) (0,1) -- (1,0);}$}%
\newcommand{\dashdotted}{\raisebox{2pt}{\tikz{$ \mathbin{ \draw[x=3ex,y=1.4ex,line width=.4ex,dashdotted, red] (0,0) -- (1,0);}$}}}%
\newcommand{\dashline}{\raisebox{2pt}{\tikz{$ \mathbin{ \draw[x=3ex,y=1.4ex,line width=.3ex,dashed,dash pattern=on 5pt off 2pt, darkgray] (0,0) -- (1,0);}$}}}%

\usepackage{cleveref}

\newcommand{\Son}{S_\mathrm{on}}
\newcommand{\Soff}{S_\mathrm{off}}

\usepackage{ulem}
\usepackage{xcolor}
\newcommand{\ADDTXT}[1]{#1}

\begin{document}

%\preprint{}

\title{Characterizing cryogenic amplifiers with a matched temperature-variable noise source} %Title of paper



\author{Slawomir Simbierowicz}
\email[Electronic mail: ]{slawomir.simbierowicz@bluefors.com}
%\thanks{}
\affiliation{Bluefors Oy, Arinatie 10, 00370 Helsinki, Finland}
\author{Visa Vesterinen}
\affiliation{VTT Technical Research Centre of Finland Ltd, QTF Center of Excellence, P.O. Box 1000, FI-02044 VTT, Finland}
\author{Joshua Milem}
\author{Aleksi Lintunen}
\author{Mika Oksanen}
\author{Leif Roschier}
\affiliation{Bluefors Oy, Arinatie 10, 00370 Helsinki, Finland}
\author{Leif Grönberg}
\affiliation{VTT Technical Research Centre of Finland Ltd, QTF Center of Excellence, P.O. Box 1000, FI-02044 VTT, Finland}
\author{Juha Hassel}
\altaffiliation{Current address: IQM Finland Oy, Keilaranta 19, 02150 Espoo, Finland}
\affiliation{VTT Technical Research Centre of Finland Ltd, QTF Center of Excellence, P.O. Box 1000, FI-02044 VTT, Finland}
\author{David Gunnarsson}
\author{Russell E. Lake}
\affiliation{Bluefors Oy, Arinatie 10, 00370 Helsinki, Finland}


\date{\today}

\begin{abstract}
We present a cryogenic microwave noise source with a characteristic impedance of 50~$\Omega$, which can be installed in a coaxial line of a cryostat. The bath temperature of the noise source is continuously variable between 0.1~K and 5~K without causing significant back-action heating on the sample space.  As a proof-of-concept experiment, we perform Y-factor measurements of an amplifier cascade that includes a traveling wave parametric amplifier and a commercial high electron mobility transistor amplifier. We observe system noise temperatures as low as $680^{+20}_{-200}$~mK at 5.7~GHz corresponding to $1.5^{+0.1}_{-0.7}$ excess photons. The system we present has immediate applications in the validation of solid-state qubit readout lines.
\end{abstract}
\maketitle %\maketitle must follow title, authors, abstract and \pacs
\section{Introduction}

\setlength{\parskip}{0pt}

Detection of weak microwave signals in a cryogenic environment is a primary task in the field of quantum technology. Solid-state qubits with transition frequencies in the microwave region operate at millikelvin bath temperatures in order to reduce thermally excited population of a qubit and to reduce noise. In practice, discriminating the state of the qubit relies on observing the qubit-state-dependent frequency shift imposed by the qubit on a coupled readout resonator \cite{blais_cavity_2004,wallraff_approaching_2005}.  The readout protocol of the microwave resonator imposes strict added-noise and gain requirements on the first amplifier in the chain to achieve single shot readout without averaging \cite{krantz_quantum_2019}. To address this challenge, various ultra-low noise amplifiers have been developed including quantum-limited parametric amplifiers exploiting the non-linearity of Josephson junctions \cite{castellanos-beltran_amplification_2008, mutus_design_2013, white_traveling_2015, macklin_nearquantum-limited_2015, zorin_traveling-wave_2017, planat_photonic-crystal_2020}, and non-linearities in superconducting thin films \cite{vissers_low-noise_2016, ranzani_kinetic_2018, chaudhuri_broadband_2017, malnou-PRXQuantum-2021}. Further development of dispersive readout requires accurate measurements of quantum-limited amplification chains within dilution refrigerators. \ADDTXT{Moreover, the study of quantum noise is directly relevant to applications in quantum state control \cite{Schoelkopf-chapter-2003}.}

Noise performance of a microwave amplifier can be expressed as a noise temperature, i.e., the noise added to a signal by an amplifier that is equivalent to that of a matched resistor of thermodynamic temperature $T$.  A common method for determining the amplifier noise temperature is the hot-cold source method or so-called Y-factor method \cite{mura_ultra-low-noise_1966,pozar_microwave_2012}. The procedure places a matched load resistor with a \ADDTXT{"hot"} temperature $T_1$ at the input of an amplifier and the amplified output noise $N_1$ is recorded using a spectrum analyzer in a specified bandwidth. The previous step is repeated using a vastly different \ADDTXT{"cold"} input noise temperature $T_2$ with corresponding amplified noise $N_2$, and the Y-factor is then defined as the ratio of the output noise powers $ Y = N_1 / N_2$. Finally, the effective noise temperature of the amplifier is obtained using the Y-factor $T = (T_1 - Y T_2) / (Y-1)$. 

Applying the Y-factor method to an amplifier in a millikelvin environment with a noise temperature in the regime $< 1$~K poses the following technical challenges. First, impedance differences between the hot and cold source signal pathways when operated through a coaxial switch can lead to systematic errors \cite{fernandez_noise-temperature_1998}. Second, as a practical challenge, cold-loads are not widely available and require users to build custom components \cite{cano_ultra-wideband_2010,yeh-jap-2017}. 
In addition, uncertainty of the two temperature points that define $Y$ can lead to fitting errors especially if the amplifier under test (AUT) has a low noise temperature compared to the cold load, requiring additional systematic correction terms \cite{otegi_uncertainty_2005}. Furthermore, if the load resistor is poorly thermalized, the calculated noise temperature from the Y-factor measurement will be overestimated \cite{cano_ultra-wideband_2010}. Other noise measurement techniques for low temperatures have been demonstrated including using a 50~$\Omega$ tunnel junction as a source of shot noise \cite{su-wei_chang_noise_2016, Spietz-science-2003, Spietz-APL-2006}. Alternatively, the power of a microwave tone can be accurately calibrated in a narrow band using the ac Stark shift of a qubit itself, and the signal to noise ratio of the transmitted tone can be measured directly \cite{macklin_nearquantum-limited_2015}. In these techniques, system complexity increases, or bandwidth decreases, which motivates manufacturable noise sources for Y-factor characterization of amplifiers with noise temperatures below 1~K.

In this article, we describe a compact noise source --- designed with a characteristic impedance of 50~$\Omega$ --- that can be operated in a temperature range between 0.1~K–-5~K within a cryostat without causing significant back action heating on the base temperature stage. As an application, we implement a typical qubit readout line \cite{krinner_engineering_2019} for circuit quantum electrodynamics\cite{blais_circuit_2020} and measure noise temperature of the amplifier cascade.  In particular, the cascade includes a traveling wave parametric amplifier (TWPA) \cite{yurke_lownoise_1996,white_traveling_2015,macklin_nearquantum-limited_2015}, a commercially available cryogenic high electron mobility transistor amplifier (HEMT) \cite{schleeh_ultralow-power_2012,gu_measurement_2013}, and additional room temperature amplification.

\begin{figure*}[t!]
\includegraphics{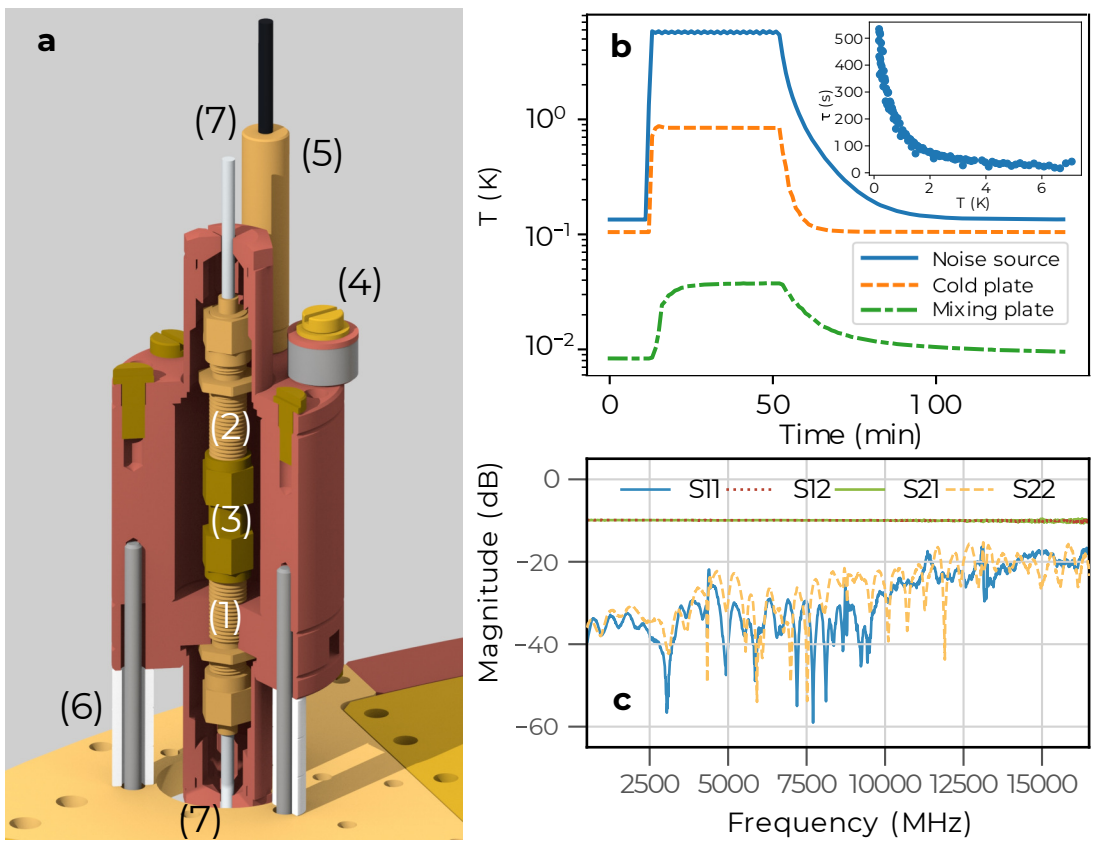}%
\caption{Matched temperature-variable noise source. \textbf{a)} Rendered cross-section of the noise source on the cold plate. The 10~dB Cu-cased bulkhead attenuators \textbf{(1)} and \textbf{(2)} are connected with an SMA adapter \textbf{(3)} and fastened to threads in the copper body at \textbf{(1)}. A heater \textbf{(4)} and thermometer \textbf{(5)} are attached on top. Thermal weak links of three stacked alumina beads \textbf{(6)} and long M3 stainless steel screws support the Cu-body and semi-rigid NbTi cables connect the microwave ports \textbf{(7)}. \textbf{b)} Temperature decay from 5~K to base. \textit{Inset:} thermal time constant as a function of source temperature. \textbf{c)} Two-port S-parameters of an attenuator calibrated at 10~mK.}%
\label{fig:source_schema}
\end{figure*}

\section{Methods}
\subsection{Thermal noise source}
We illustrate the noise source and its characteristics in Fig.~\ref{fig:source_schema} and a photograph of its parts is provided in Appendix~\ref{appb}. The central elements are the matched resistor networks that we implement using two 10~dB bulkhead copper-cased attenuators with K-2.92mm jack interfaces shown as (1) and (2) in Fig.~\ref{fig:source_schema}a. \ADDTXT{The attenuator is realized with an on-chip Ni--Cr thin film resistor network that is clamped within a Cu (C101 alloy \cite{finger-ijimw-2008}) case. The inner contacts of the attenuator are made from gold-plated Be--Cu alloy.} The thermalization of the copper-cased attenuator is more efficient than that of a commercial stainless steel counterpart and one of the bulkheads (1) is fastened directly to threads in the copper body of the noise source. The operation principle of the noise source is that the resistive heater bobbin of (4) increases the temperature of the Cu body (with a mass of approximately 0.45~kg) including the inner bulkhead attenuators of (1) and (2) to emit thermal noise into the coaxial line. Temperature monitoring and feedback are implemented using a ruthenium oxide thermometer attached to the body of the noise source in (5). The thermometer is calibrated against a noise thermometer (a magnetic field fluctuation thermometer, from Magnicon GmbH) which was calibrated at Physikalisch-Technische Bundesanstalt against the PLTS-2000 \cite{engert_low-temperature_2013,engert_new_2016}. The source can be operated either by transmitting a microwave probe tone through (1) and (2), or as a single-ended noise source where noise is emitted directly from one of the ports. The source is designed to be mounted to an ISO-KF\ADDTXT{25} port in a dilution refrigerator flange so that a superconducting semi-rigid coaxial cable can pass through the port opening as shown in Fig.~\ref{fig:source_schema}a. In the measurements presented in this article, the noise source is mounted to a stage with a base temperature of 100~mK, i.e., the cold-plate stage of a commercial dilution refrigerator cryostat, where the mixing chamber stage below it reaches a temperature of approximately 8~mK.

To limit unwanted heating of the surrounding areas of the cryostat, the three attachment points of (6) are designed as thermal weak links that consist of three M3x50~mm stainless steel screws running through alumina fish-spine beads with an area-to-length ratio of 3.9~mm. At high temperatures of greater than 3~K, conduction through the alumina beads dominates the heat transfer so that the noise source can be cooled quickly during the initial cooldown. However, at lower temperatures, thermal conductance of the alumina beads becomes weak compared to stainless steel screws, and the screws are the thermal weak link between the noise source and flange. For example, if the noise source is at the temperature of 1~K, and the supporting flange is 100~mK, the heat conductivity integrals \cite{pobell_matter_2007} are evaluated as $\dot{Q}_{\textrm{S}} = 7.2 \times 10^{-4}$~W~/~cm and $\dot{Q}_{\textrm{AlOx}} = 7.8 \times 10^{-5}$~W~/~cm for the stainless steel and alumina beads respectively. Taken together, the total heating power on the mounting flange due to the noise source is 100~$\mu$W. Based on simple estimations, any residual heating due to radiation is at least three magnitudes lower even if the noise source were at 5~K. Furthermore, connections between the attenuator ports and the other stages of the cryostat are made with thin superconducting NbTi cables to minimize heat transfer through the microwave cabling. \ADDTXT{The coaxial cables directly connected to the noise source have an NbTi center conductor diameter of 0.2~mm, PTFE dielectric outer diameter of 0.66~mm, and NbTi outer conductor diameter of 0.9~mm. Non-magnetic gold-plated Be--Cu connectors are soldered to the NbTi coaxial cable ends using standard techniques such as those in Ref.~\citenum {debry-rsi-2018}. The impedance specification for this cable is $(50  \pm 3)$~$\Omega$ at 4~K.}
 
We studied the back-action heating of the noise source on its surroundings and display the results in Fig.~\ref{fig:source_schema}b. In practice, we find that the source can be operated between 0.1~K and 5~K when placed on the cold plate of a dilution refrigerator without significant heating of the lowest temperature stage (mixing chamber plate). Importantly, this provides enough dynamic range in output noise temperature to analyze both near-quantum limited-amplifiers and low noise HEMT amplifiers. In Figure~\ref{fig:source_schema}b, at a noise source temperature of 5~K the mixing chamber heats up to 30~mK. This unwanted heating of the sample space could be further reduced (at the expense of a longer time constant) by increasing the length of the stainless steel screws and number of alumina beads. The thermal time constant of the noise source is temperature-dependent and governed by the ratio of the heat capacity of the copper-body of the noise source to the thermal conductance of the weak link \cite{bachmann_heat_1972}. We measure the time constant by first elevating the noise source temperature to a constant value using a constant heater power.  Subsequently, the heating power is reduced by 5~\%, and the cooling curve between each steady state temperature was fit to an exponential decay with time constant~$\tau$. The procedure was repeated over a range of heater powers that correspond to the temperature points in the inset of Fig.~\ref{fig:source_schema}b. As the temperature increases from 0.1~K to 5~K, the corresponding thermal time constant of the noise source decreases from > 500~s to 10~s. In the following sections, waiting times of at least $>2.5 \tau$ between temperature data points are used. For example, to guarantee thermalization of the noise source in the results of Sec.~\ref{sec:results}, we always wait 20~minutes after changing the temperature feedback setpoint and before starting a measurement. 

We characterize the impedance matching of the same type of attenuator as the ones installed in the noise source by measuring the calibrated reflection and transmission using a \ADDTXT{multiline} thru-reflect-line technique \ADDTXT{originally outlined in Refs.~\citenum{ranzani_two-port_2013,yeh-rsi-2013} and further described in Appendix~\ref{appa}.} As shown in Fig.~\ref{fig:source_schema}c, the reflected signals $S_{11,22}$ are below -20~dB over a frequency span from \ADDTXT{500~MHz -- 11~GHz} at the temperature of 10~mK. Based on this impedance matching, we omit mismatch effects in the noise error budget of the following sections. Furthermore, we observe that the attenuation remains at a nearly constant value of \ADDTXT{9.977~dB} over the measurable frequency range \ADDTXT{and that its deviation from the room temperature attenuation value is negligible ($< 0.02\mbox{ }\%$ change)}. The bandwidth of the measurement is limited by the directional coupler in the measurement.

\subsection{Measurement Principle}\label{sec:principle}
\begin{figure}[b!]
\includegraphics[width=1.0\linewidth]{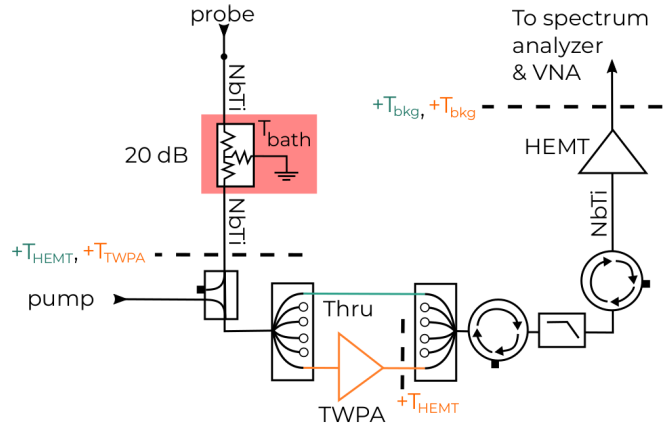}%
\caption{Noise source wired with microwave switches for comparing noise and gain with and without a TWPA. A probe tone is thermalized to the mixing chamber stage temperature using attenuators before being routed to the input of the noise source at $T_\mathrm{bath}$. The first signal path includes a coaxial "Thru" at the mixing chamber stage to bypass the TWPA and enable loss-calibrated gain measurements. The second signal path includes the microwave-pumped TWPA to amplify both the probe tone and noise. Subsequent isolators protect the TWPA from HEMT back-action. The green (orange) variables indicate lumped noise contributions for the first (second) signal path. 
}
\label{fig:cascade_schema}
\end{figure}

Figure~\ref{fig:cascade_schema} displays the noise source with controllable physical temperature $T_{\textrm{bath}} \geq 100$~mK installed at the input of the cascaded amplifiers. A full wiring diagram with all components is shown in Appendix~\ref{appa1} (Fig.~\ref{fig:full_twpa_fig}a). The physical temperature determines the input noise temperature $T_\mathrm{in}$ in either a classical or quantum regime \cite{clerk_introduction_2010}. In the classical regime of $k T_{\textrm{bath}} \gg hf$ (where $k$ is the Boltzmann constant, $h$ is the Planck constant, and $f$ is the operation frequency of the AUT) we consider the condition $T_{\mathrm{in}} \approx T_{\mathrm{bath}}$ \cite{pozar_microwave_2012}.  In contrast, the quantum regime is characterized by $k T_{\textrm{bath}} \lessapprox hf$, with a correspondence between $T_\textrm{bath}$ and $T_\textrm{in}$ as,
\begin{equation}\label{eq:input_noise}
T_\mathrm{in} = \frac{hf}{2k}\coth{\frac{hf}{2kT_\mathrm{bath}}} \textrm{.}
\end{equation}
To improve accuracy, we always include this quantum correction to the noise emitted by the noise source, even at higher noise temperatures. 

%\clearpage

The wiring configuration of Fig.~\ref{fig:cascade_schema} is designed to compare noise results with and without the TWPA. Specifically, two distinct signal paths can be selected with the coaxial switch that is installed at the millikelvin stage. The first path includes only a 76-mm-long coaxial transmission line labeled "Thru" (green line) to bypass the TWPA, and the second signal path includes the TWPA (orange triangle). The microwave pump tone required to operate the TWPA can be combined with the signal paths through a directional coupler at the mixing chamber stage of the cryostat.  After the HEMT and room temperature amplification (not shown \ADDTXT{in Fig.~\ref{fig:cascade_schema}}), a spectrum analyzer detects the output noise power from the system. In particular, we measure the integrated noise power $P_\mathrm{out}$ from a 10~kHz \ADDTXT{span} \ADDTXT{to represent noise} at a single operating frequency. \ADDTXT{Additionally, we apply a probe tone at the center of the selected span. This enables accurate detection of the difference in magnitudes of transmission between the first and second signal paths to determine the gain of the TWPA. We set the resolution bandwidth of the spectrum analyzer to 100~Hz in order to keep the incoming noise at least 10~dB above the noise floor of the spectrum analyzer as well as to detect the weak probe at a signal-to-noise ratio of at least 10~dB. Since the signal-to-noise ratio allows it, the probe power is set to a safe low value of approximately -130~dBm at the input of the noise source, both to prevent pre-amplifier saturation and to limit self-heating of the noise source.}

%Specifically, we measure the integrated noise power $P_\mathrm{out}$ from a 10~kHz window at a single operating %frequency. 

%\ADDTXT{Specifically, we measure the integrated noise power $P_\mathrm{out}$ from a 10~kHz window at a single operating frequency. Additionally, we apply a probe tone at the center of the selected window. This allows to accurately detect the difference in magnitude of transmission between the first and second signal paths to determine the gain of the TWPA.
%Then we set the resolution bandwidth of the spectrum analyzer to 100 Hz, in order to keep the incoming noise at least 10 dB above the noise floor of the spectrum analyzer as well to detect the weak probe at a signal-to-noise ratio of at least 10~dB. Since the signal-to-noise ratio allows it, the probe power is set to a safe low value of approximately -130 dBm at the input of the noise source, both to counter pre-amplifier saturation and to limit self-heating of the noise source.
%}

We consider initially the output power of the first signal path that bypasses the TWPA. To express the measured output noise power in terms of equivalent noise temperatures of individual components, we divide $P_\mathrm{out}$ by the sum of amplifier gain and component losses $G_\mathrm{tot}$, the measurement bandwidth $B$\ADDTXT{~$=100$~Hz}, and $k$ as
\begin{equation}\label{eq:noise_hemt}
\frac{P_\mathrm{out}}{G_\mathrm{tot} k B} =  T_\mathrm{in} +T_\mathrm{HEMT} + \frac{T_\mathrm{bkg}}{G_\mathrm{HEMT}} \mathrm{,}
\end{equation}
where $T_\mathrm{in}$ is the experimentally controllable input noise temperature from the noise source via Eq.~\eqref{eq:input_noise}, $T_\mathrm{HEMT}$ is the noise temperature of the HEMT with gain $G_\mathrm{HEMT} \sim 10^{4}$, and $T_\mathrm{bkg}$ arises from the amplification and loss in components at higher temperature stages following the HEMT. The green text labels ($T_{\textrm{HEMT}}$,$T_{\textrm{bkg}}$) in Fig.~\ref{fig:cascade_schema} indicate the noise of lumped noise contributions along the first signal path. These contributions govern the system noise $T_\mathrm{sys} = P_\mathrm{out}  / G_\mathrm{tot} k B$, which is the key quantity when determining the performance of the amplifier chain.

We collect many pairs of data points $(T_{\mathrm{in}}$, $P_{\textrm{out}})$ for $0.135 \mbox{ K} <T_{\textrm{bath}} < 3.6 \mbox{ K}$ and fit these with Eq.~(\ref{eq:noise_hemt}) reformulated as $T_{\textrm{in}} = \alpha P_{\textrm{out}} - T_\mathrm{added}$, which we demonstrate with an example fit in Appendix~\ref{appa2}. This reordering of terms is done because the gain $G_\mathrm{tot}$ contained in the slope $\alpha$ cannot be measured directly with the wiring of Figs.~\ref{fig:cascade_schema} and \ref{fig:full_twpa_fig}. The offset $T_\mathrm{added}$ reveals the noise added by the cascade, which is equivalent to the system noise after the addition of a minimum input noise $T_\mathrm{bath} = 10\mbox{ mK}$ using Eq.~\eqref{eq:input_noise}. In other words, nearly $hf/2$ input noise would emerge as the noise coming from the 10-mK mixing chamber stage itself, even in the absence of experimentally added noise. In addition to amplification, the measured system noise quantifies the noise contributions due to the loss in passive components such as the cables, isolators, and filters between the noise source and HEMT. On the other hand, the high gain of the HEMT makes noise after the HEMT insignificant given the condition $T_\mathrm{bkg} / G_\mathrm{HEMT} \ll T_\mathrm{HEMT}$.

For the second signal path of Fig.~\ref{fig:cascade_schema} that includes the TWPA, we expect to obtain a lower system noise temperature due to the added gain from the TWPA as well as its low physical temperature and composition from low-loss superconducting materials. More precisely, the TWPA is an approximately 50-ohm transmission line composed of a series array of 4064 Josephson junctions with a critical current of about 4~$\mu$A, fabricated at VTT with the side-wall spacer passivated junction process \cite{gronberg_side-wall_2017}. Additional details including a list of design parameters, an electrical circuit model, and a microphotograph of the device are available in Appendix~\ref{appc1}. Using a strong pump tone, the Kerr-nonlinearity of the junctions allows a four-wave mixing process where an additional idler appears at $f_\mathrm{i} = 2f_\mathrm{p}-f_\mathrm{s}$ with $f_\mathrm{s}$ as the signal and $f_\mathrm{p}$ the pump frequency. This process adds additional steps to the measurements and data analysis since the idler band is also a source of noise, as we detail in Appendix \ref{appa2}. Another consideration is that, to prevent saturation of the TWPA, we operate the noise source at low temperatures of $T_{\textrm{bath}} < 900 \mbox{ mK}$. In this low temperature regime, the quantum corrections of Eq.~\eqref{eq:input_noise} are especially important \cite{clerk_introduction_2010}. 

\ADDTXT{The wiring schematic of Fig.~\ref{fig:cascade_schema} enables the operation of the TWPA with negligible perturbation on the noise source. The pump tone (of approximately 0.14~nW of microwave power at the TWPA input) is introduced through a dedicated line and is combined with the signal path immediately before the TWPA at the base stage of the cryostat with the directional coupler shown in Fig.~\ref{fig:cascade_schema}. This directional coupler isolates the noise source from the incoming pump by a directivity factor of 15~dB. In addition, the power reflected from the TWPA back to the noise source is suppressed by 16.5~dB with respect to the incoming pump given the reflection coefficient of Table~\ref{tab:twpa_table} in Appendix~\ref{appc}. Therefore the pump power reaching the noise source is seven orders of magnitude lower than the heat applied directly to the noise source at the noise source temperature of $T_{\textrm{bath}} = 1$~K.}

\section{Measurement Results and Analysis}\label{sec:results}
\begin{figure}[t]
\includegraphics[width=1.0\linewidth]{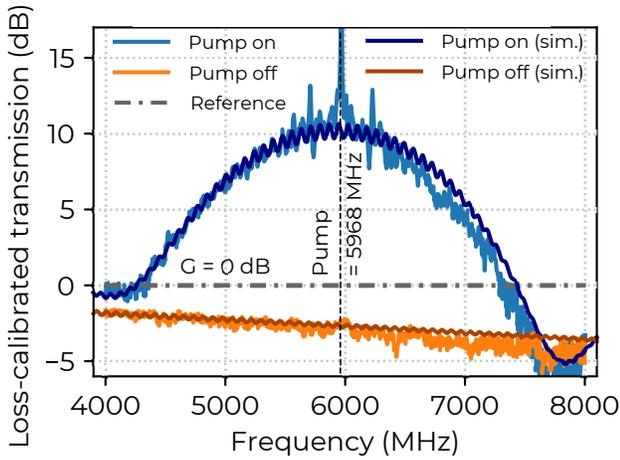}%
\caption{Traveling wave parametric amplifier gain (loss) in blue (orange) overlaid with navy-colored (brown) numerical simulation data (see appendix \ref{appc}). The vertical line marks the pump frequency.}%
\label{fig:twpa_fig}
\end{figure}

We characterize the TWPA gain using the wiring introduced in Sec.~\ref{sec:principle} and display the results in Fig.~\ref{fig:twpa_fig} measured with a network analyzer. The device is pumped at 5968~MHz with an approximate power of -69~dBm at its input port (that was estimated by accounting for $\sim 10$~dB loss arising from cable losses). Using a probe tone of approximately -131~dBm at the TWPA input port, we observe the characteristic semi-circle gain profile that is centered on the pump frequency as can be seen in the blue "pump on" trace of Fig.~\ref{fig:twpa_fig}. In comparison, when the pump is turned off we observe attenuation from dielectric loss in the transmission line of the TWPA (orange trace). The loss (4.5~dB at 8~GHz) in the un-pumped device motivates using the separate coaxial "Thru" for accurate loss-calibrated gain measurements of parametric amplifiers that are operated in transmission. The calibrated gain of the TWPA stays above 10~dB in a 700~MHz region around the pump.

After verifying the TWPA operation we are now prepared to perform the Y-factor measurements for both signal paths following the measurement principle of Sec.~\ref{sec:principle} using many input noise temperature points. We plot the noise temperature results in Fig.~\ref{fig:noise_twpa} where the error bars are derived in Appendix \ref{appa3}. For the "Thru" signal path that bypasses the TWPA, the noise is dominated by the wiring and connectors between the noise source and HEMT, filters, isolators, and the added noise of the HEMT. The system noise---plotted as blue markers of Fig.~\ref{fig:noise_twpa}---is measured at frequencies between 5~GHz -- 7~GHz, and its lowest value is $3.3^{+0.1}_{-0.8}$~K near 6~GHz.  The noise exceeds the values reported in the data sheet of the HEMT (dashed line) of approximately 1.7 -- 1.9~K. We can explain this discrepancy by performing a separate control experiment where we remove the isolators, and filters from the output line. In the control experiment, we operate the noise source in a single-ended mode at 6.5~GHz (red cross in Fig.~\ref{fig:noise_twpa}) with the source directly connected to the HEMT through coaxial cables (see schematic of Appendix~\ref{appa1}, Fig.~\ref{fig:full_twpa_fig}b). The system noise temperature in the control experiment is 2.2$_{-0.3}^{+0.1}$~K which is in agreement with the value reported in the HEMT data sheet. Therefore, we conclude that the wiring, isolators, and filters have a non-negligible contribution to the system noise. Regardless, in any practical quantum measurement, one needs to isolate a sample or a parametric amplifier from a reflective and relatively high temperature HEMT. Therefore, the higher-than-specified system noise values represent realistic system performance.

\begin{figure}
    \vspace{-1.1cm}
    \centering
    \includegraphics[width=1.0\linewidth]{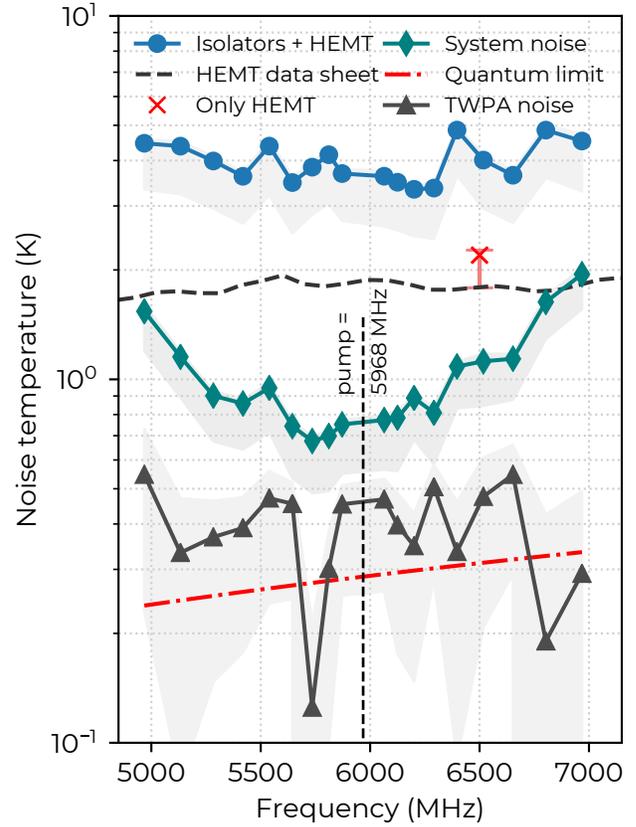}
    \vspace{-0.7cm}
    \caption{Noise temperatures given by the Y-factor method. The system noise is measured with (\protect\markertwo) and without the TWPA (\protect\markerone), which allows calculating the intrinsic TWPA noise (\protect\markerthree) bounded by the standard quantum limit (\protect\dashdotted). \ADDTXT{The solid lines guide the eye and the} vertical line marks the pump at 5968~MHz.
    Measuring the HEMT separately without isolators (\protect\Cross) allows comparison to data sheet values (\protect\dashline). \ADDTXT{The shaded regions represent errors from} insertion loss, and fitting uncertainty.}%
    \label{fig:noise_twpa}
\end{figure}

Measured through the signal path that includes the TWPA, the system noise (teal diamonds) reaches its lowest value of $680^{+20}_{-200}$~mK at a frequency of 5735~MHz (which is detuned by 233~MHz from the pump). This result is $2.5^{+0.1}_{-0.7}$ times the one-photon quantum limit \cite{caves_quantum_1982} (red dash-dotted line) that also includes input noise. We note that the measured system noise follows an inverted trend with respect to the frequency-dependent gain profile of Fig.~\ref{fig:twpa_fig}, because the gain of the low-noise TWPA can dilute the added noise of the HEMT in the total system noise.

Finally, we calculate the intrinsic noise of the TWPA by analyzing the differences in system noise and gain between the two signal paths. We reduce the system noise from the "Thru" signal path to the input of the TWPA and calculate the difference which is the TWPA noise $T_\textrm{TWPA} = T_\textrm{sys,TWPA} - T_{\textrm{sys,HEMT}}/ G_\textrm{TWPA}$. \ADDTXT{The relatively higher scatter and wider error bars (shaded region) on this derived quantity is the result of the error in the system noises from the two paths adding together, since the gain of the TWPA is not enough to completely eliminate the contribution of the HEMT to the total system noise. In the future, this error could be decreased by increasing the resolution bandwidth accompanied by a moderate improvement to pre-amplification, which would make the measurement faster and reduce statistical uncertainty in the acquired noise. However, we find that the strong pump easily saturates amplifiers that follow the TWPA and hence we were unsuccessful in adding more pre-amplification for this experiment. Without additional pre-amplification, the resolution bandwidth could still be increased to approximately 1~kHz by rewiring the coupler at room temperature to use the through line for the noise acquisition.}

Another representation \ADDTXT{for the noise of the TWPA} can be obtained by converting the \ADDTXT{intrinsic} noise temperature to photons. We calculate the weighted average of input noise photons with $n = k T_\textrm{{TWPA}} / h f $, leading to an estimate of \ADDTXT{$\bar{n} = 1.3^{+0.3}_{-0.2}$} photons, which is close to the one-photon quantum limit.  This analysis supports the hypothesis that if the amplifier gain increased, the system noise would asymptotically decrease toward the quantum limit to achieve optimal signal-to-noise.

\section{Conclusion}

In conclusion, we have developed a well-matched temperature controllable "cold" noise source designed for characterization of ultra-low noise amplifiers in a cryogenic environment. The dynamic range is designed for modern cryogenic amplifiers including both quantum limited parametric amplifiers and HEMTs.  We demonstrate its use in conjunction with an amplifier cascade intended for qubit readout. Our results also highlight the significance of the losses in components that directly precede the HEMT, which we observe can even double the system noise temperatures. The noise source apparatus lays the groundwork for testing other sensitive low-noise amplifiers and optimized readout line configurations. 

%\vspace{-2cm}
\begin{acknowledgments}
S.S. thanks E.T. Mannila for useful discussions on noise measurements. V.V., L.G., and J.H. acknowledge financial support from the Academy of Finland through its Centers of Excellence Program (project nos. 312059 and 312294) and grant no. 321700, and from the EU Flagship on Quantum Technology H2020-FETFLAG-2018-03 project 820363 OpenSuperQ. Furthermore, the authors gratefully acknowledge Janne Lehtinen and Mika Prunnila from VTT Technical Research Center of Finland Ltd also involved in the TWPA
development. \ADDTXT{This research made use of scikit-rf, an open-source Python package for RF and Microwave applications.}
%\vspace{-1cm}
\end{acknowledgments}

%\clearpage
%\newpage
\appendix
%\newpage

\section{Photograph of noise source}\label{appb}

\begin{figure}[hbt]
%\vspace{-0.7cm}
\includegraphics[width=3.37in]{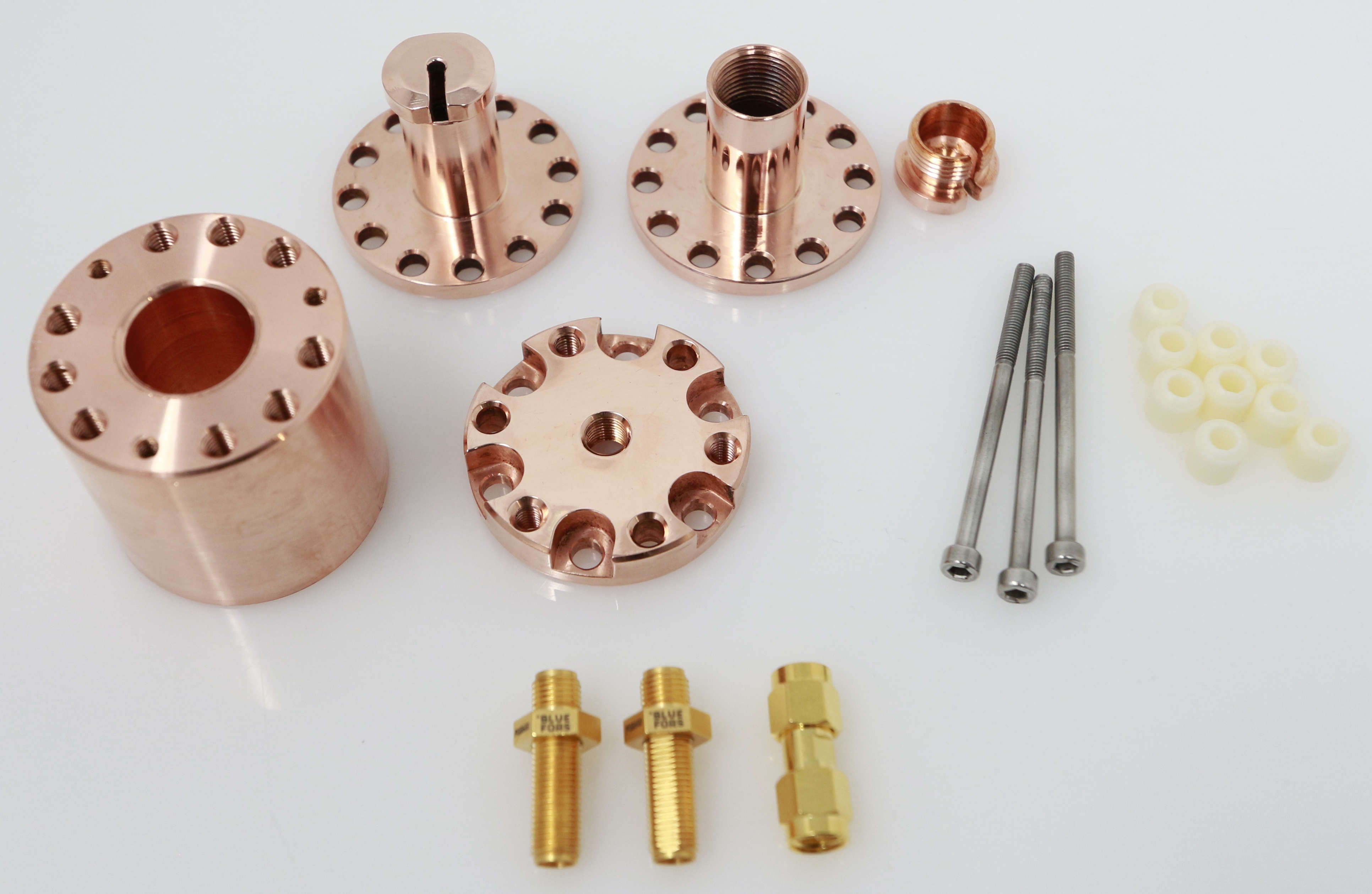}
\caption{Photograph of the parts used for building the noise source.}
%\vspace{-0.3cm}
\label{fig:source_photo}
\end{figure}

Photograph of the parts used for building the noise source (Fig.~\ref{fig:source_photo}).

\section{Details on measurement and analysis}\label{appa}

%\vspace{-0.2cm}
\subsection{Full Measurement Schematic}

\begin{figure}[hbt]
    \centering
    %\vspace{-0.7cm}
    \includegraphics[width=1\linewidth]{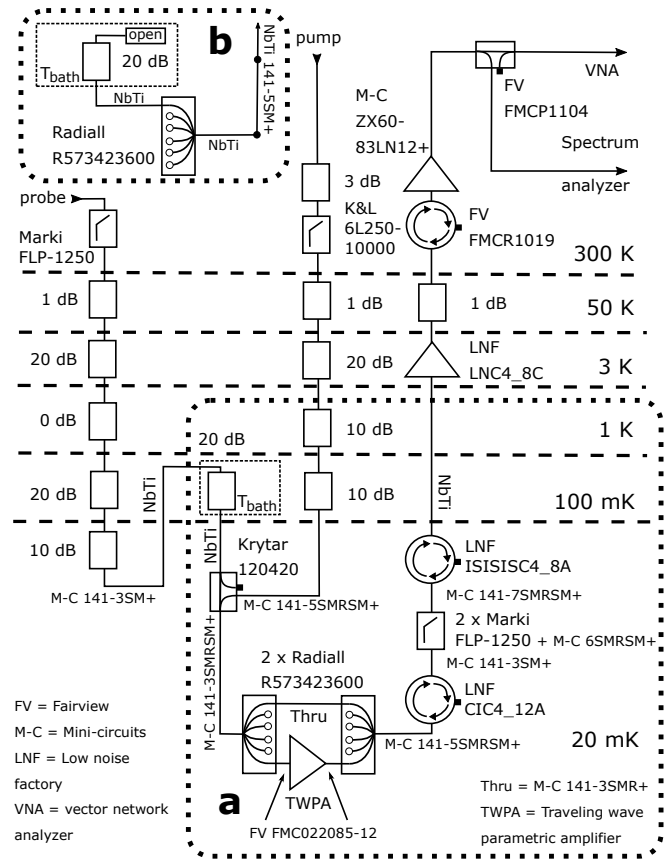}
    \caption{Detailed schematic for noise and gain measurements. \textbf{a)} The probe signal is filtered and attenuated in stages with 71~dB of fixed attenuation. The noise source is weakly thermalized on the cold plate (100~mK). The TWPA pump undergoes 64 dB of fixed attenuation before reaching the device between microwave switches. The amplified signal is measured on the spectrum analyzer (Signalhound USB-SA124B) or the VNA (N5232B). \textbf{b)} Control experiment with the noise source connected to the HEMT without isolators between them. The input port is left open. \ADDTXT{The electromechanical switches (R573423600) in both panels are single pole six throw latching switches.}}
    %\vspace{-0.6cm}
    \label{fig:full_twpa_fig}
\end{figure}

The detailed wiring schematic for noise and gain measurements is shown in Fig.~\ref{fig:full_twpa_fig}.

\label{appa1}
%\newpage

\clearpage
\subsection{Data Analysis}
\label{appa2}

In addition to the quantum corrections presented in Eq.~\eqref{eq:input_noise} of the main text, when considering the added noise due to the parametric processes in the TWPA, the presence of the additional idler tone needs to be taken into account. This can be done by introducing an effective input noise as
\begin{equation}\label{eq:noise_eff}
\begin{split}
    & T_\mathrm{in,eff} = T_\mathrm{in} + \frac{G_\mathrm{conv}}{G_\mathrm{TWPA}}T_\mathrm{idler} \\ & =  \frac{hf_\mathrm{s}}{2k}\coth{\frac{hf_\mathrm{s}}{2kT_\mathrm{bath}}} + \frac{G_\mathrm{conv}}{G_\mathrm{TWPA}}\frac{hf_\mathrm{i}}{2k}\coth{\frac{hf_\mathrm{i}}{2kT_\mathrm{bath}}} \textrm{,}
    \end{split}
\end{equation}
where the first term constitutes noise from the signal frequency $f_\mathrm{s}$. The second term, arising from four wave mixing, is noise from the idler frequency $f_\mathrm{i} = 2f_\mathrm{p}-f_\mathrm{s}$ converted at a gain of $G_\mathrm{conv}$ and reduced back to the input by division with the gain of the TWPA $G_\mathrm{TWPA}$. Now, using the effective input noise $T_\mathrm{in,eff}$, we may write the system noise equation for the path of Fig. \ref{fig:cascade_schema} that includes the TWPA in a similar manner as was done for the "Thru" path in the main text. This time we refer to the orange variables and write the system noise as
\begin{equation}\label{eq:noise_twpa}
    T_\mathrm{sys,TWPA} = \frac{T_\mathrm{bkg}}{G_\mathrm{HEMT}G_\mathrm{TWPA}}+\frac{T_\mathrm{HEMT}}{G_\mathrm{TWPA}}+T_\mathrm{TWPA}+T_\mathrm{in,eff} \mathrm{,}
\end{equation}
where $T_\mathrm{TWPA}$ is the intrinsic noise of the TWPA, $T_\mathrm{bkg}$ is the same background noise from Eq.~\eqref{eq:noise_hemt} again made negligible by HEMT gain $G_\mathrm{HEMT}$, whereas the HEMT noise $T_\mathrm{HEMT}$ is approximately equal to that measured using the "Thru" path, since the lumped contributions differ by only a few low-loss components. Therefore, measuring the system noise through both paths allows calculating the intrinsic noise of the parametric amplifier. The error arising from the differing components is taken into account following the principles presented in the next subsection.

In order to apply the above equations, the noise measurements are performed in a similar manner, as for the "Thru" path of Fig. \ref{fig:cascade_schema}: we heat the source, wait for 20~minutes for the proportional integral differential (PID) feedback to stabilize the temperature, and measure spectra in 10~kHz windows centered on the chosen signal frequencies. A slight difference in the TWPA measurements is that the spectra are also captured at the corresponding idler frequencies to determine signal and conversion gains accurately. This procedure is repeated over a range of bath temperatures $T_\mathrm{bath}$ between 0.135 and 0.9~K.

\begin{figure}[ht]
\includegraphics{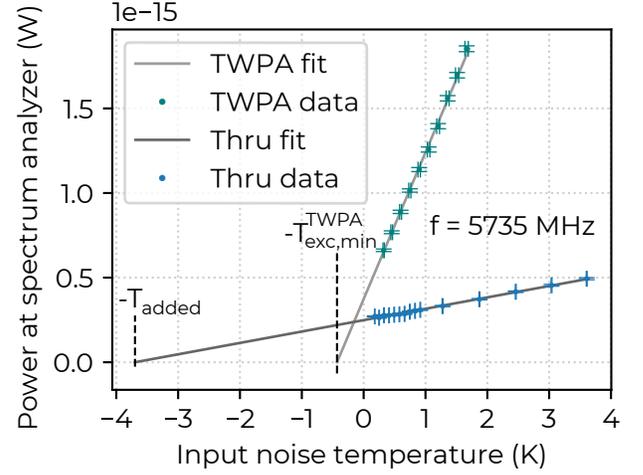}
\caption{Example Y-factor fits at optimal TWPA working point of 5735~MHz. Noise power measured from the "Thru" (TWPA) path versus input (effective input) noise is shown in blue (teal). Fits in dark (light) gray give the added (excess) noise offsets marked by vertical dashed lines. Error bars consist of calibration uncertainties and for the TWPA also error in gain measurement.}
\label{fig:yfactor_example}
\end{figure}

Using Eqs.~\eqref{eq:input_noise}, \eqref{eq:noise_eff}, and \eqref{eq:noise_twpa}, we then fit linear equations of the form $ T_\mathrm{in,eff}=\alpha P_\mathrm{out}-T_\mathrm{exc,min}^\mathrm{TWPA}$ where $\alpha$ is proportional to the total gain and $T_\mathrm{exc,min}^\mathrm{TWPA}$ is the excess noise over the one-photon quantum limit for system noise, when only vacuum noise is assumed at the input. An example fit, corresponding to the lowest measured system noise of $680^{+20}_{-200}$~mK at 5735~MHz is shown in Fig.~\ref{fig:yfactor_example} alongside a corresponding fit for the "Thru" path as per Eq.~\eqref{eq:noise_hemt}. Now, evaluating $T_\mathrm{exc,min}^\mathrm{TWPA}+T_\mathrm{in,eff}$ at $T_\mathrm{bath}=10 \text{ mK}$ gives us the lowest system noise temperature we could realistically obtain if the amplifier cascade were wired to receive input noise from the mixing chamber stage at base temperature. The fit is weighted using calibration errors of the thermometer and the spectrum analyzer which we estimate to be in the order of 5~mK--7~mK and 0.25~dB, respectively. Actuating the rf switch heats up the mixing chamber, and therefore we only toggle between probe paths after measuring all the temperature points for a path. For this reason, the effective input noise $T_\mathrm{in,eff}$ of Eq.~\eqref{eq:noise_eff} is determined using averaged signal and conversion gains to even out any drifts and their errors appear also in horizontal uncertainty of the teal-colored data of Fig.~\ref{fig:yfactor_example}, as we explain in the upcoming subsection.

%\newpage

% If in two-column mode, this environment will change to single-column format so that long equations can be displayed. 
%\begin{widetext}
%$$\mbox{put long equation here}$$
%\end{widetext}

% Figures should be put into the text as floats. 
% Use the graphics or graphicx packages (distributed with LaTeX2e).
% Put the label that you will use with \ref{} command in the braces of the \label{} command.
%

%\newpage

\subsection{Error analysis}\label{appa3}
In our noise calculations shown so far, we have omitted the microwave circuitry after the HEMT as well as the insertion loss from rf switches, filters, isolators, and other wiring between the noise source and the first amplifier. In reality, to obtain the error bars for instance for the noise of "Thru" path in Fig.~\ref{fig:noise_twpa}, we have already corrected the system noise of Eq.~\eqref{eq:noise_hemt} with
\begin{equation}\label{eq:noise_hemt_detailed}
    T_\mathrm{sys,HEMT} = -G_\mathrm{ATT}T_\mathrm{added}-\frac{T_\mathrm{bkg}}{G_\mathrm{HEMT}}-G_\mathrm{ATT}T_\mathrm{ATT} + T_\mathrm{in} \mathrm{,}
\end{equation}
where $T_\mathrm{added}$ is again the offset from the fit, $G_\mathrm{ATT}$ and $T_\mathrm{ATT}$ are the loss and temperature of the attenuator modeling the insertion loss. The input noise of 10~mK does not affect the error estimation:
\begin{equation}\label{eq:noise_hemt_error}
\begin{split}
    & dT_\mathrm{HEMT} = |-G_\mathrm{ATT}|dT_\mathrm{added}+|-T_\mathrm{added}|dG_\mathrm{ATT} \\
    & + \left|-\frac{1}{G_\mathrm{HEMT}}\right|dT_\mathrm{bkg}+\left|\frac{T_\mathrm{bkg}}{G_\mathrm{HEMT}^2}\right|dG_\mathrm{HEMT} \\
    & +|-G_\mathrm{ATT}|dT_\mathrm{ATT}+|-T_\mathrm{ATT}|dG_\mathrm{ATT} \mathrm{,}
    \end{split}
\end{equation}
where the third and fourth terms are still very small due to the high gain of the HEMT, $dT_\mathrm{added}$ is the error of mean given by least squares, $G_\mathrm{ATT} = 1$ is the uncorrected insertion loss, and $dT_\mathrm{ATT}$ and $dG_\mathrm{ATT}$ are the corrections to switch temperature and insertion loss. We feel that the latter errors can only reduce the noise temperature and they should equate to zero for positive errors. In the negative case, reasonable estimates are -100~mK and the difference of 0~dB and the maximal estimated loss of -1~dB. The latter number is based on data sheet values for the rf switches and the isolators before the HEMT amplifier.

The error in system noise temperature for the path that includes the TWPA, i.e. Eq.~\eqref{eq:noise_twpa}, is derived quite similarly but we may now limit the insertion loss to 0.5~dB from the first rf switch. A new consideration is that, the variations of the probe $G_\mathrm{TWPA}$ and conversion gains $G_\mathrm{conv}$ also have to be taken into account in the effective input noise of Eq.~\eqref{eq:noise_eff}, as it is needed in fitting and its error defines the respective weights for individual data points. Since we always operate at a signal-to-noise ratio of over 10~dB, even at the highest source temperature, noise power at the probe frequency should not affect the observed magnitude of the probe signal on the spectrum analyzer. Therefore, for both of these types of gains, at a given frequency, we may take the mean value (of all gains of the same type) across all source temperatures. We then use these means and their statistical errors in calculating the effective input noise and its uncertainty for that frequency. 

\ADDTXT{Using the above method, we avoid switching during the sweep through the temperature operating points, and therefore switching is done only once during the whole experiment. A switching event heats up the mixing chamber slightly and for this reason we monitor the mixing chamber temperature and wait for it to cool below 20~mK before we start the measurement through the other path. So any error due to repeatibility of switch insertion loss will be systematic in all the noise power data measured for a single path, which means that it only affects the gain values obtained and therefore the estimate for intrinsic noise of the TWPA. We find that the errors in the latter are more influenced by the fact that it is a difference of two quantities, as we also mention in the main text, and by the error in gains which are of the order of 1~dB. To investigate the latter point, we tested the switch repeatability in a separate cooldown. With shorting caps on each switch port on nominally identical switches, we cycled through all the switch channels ten times at base temperature, with a waiting time of 2~minutes between switching events. In this way, we found that the variability indicative of insertion loss is less than 0.5~dB from 6 to 7~GHz and negligible below 6~GHz.}

\subsection{Cryogenic S-parameter measurement}\label{appa4}
\ADDTXT{A multiline Thru-Reflect-Line (TRL) calibration procedure was performed to measure cryogenic S-parameter data of the attenuator used in the noise source by implementing the methods previously outlined in Refs.~\citenum{ranzani_two-port_2013, yeh-rsi-2013}. The two input and two output signals were sourced and received by a pre-calibrated four-port vector network analyzer. The signals were routed through directional couplers at the MXC stage and into single-pole six-throw latching electromechanical switches so that transmission and reflection could be acquired from both ports of a two-port device under test connected between the switches.  Thru, Reflect, Line~1, Line~2, and the 10~dB attenuator-under-test were connected with identical length cables (phase-matched within <1~ps) and connected between the switch ports. The Thru, Line~1, and Line~2 were comprised of sections of rigid coaxial transmission line with different physical lengths of 22.23~mm, 25.58~mm, and 46.66~mm respectively. The coaxial transmission lines were designed with a nominal impedance of 50~$\Omega$ with an inner contact diameter of 1.27~mm, and a Teflon-dielectric outer diameter of 4~mm. These components were manufactured with cryogenic compatible metals: gold-plated C101 Cu alloy housing and a Be--Cu gold-plated center conductor. The Reflect standards were realized with a shorting cap that shunted the center conductor to its shield. The calibration algorithm was implemented with an open-source software tools in \url{www.scikit-rf.org}.}

\ADDTXT{The Thru, Line 1, and Line 2 standards were pre-characterized at room temperature using a commercial network analyzer with an electronic calibration kit using a reference input impedance of 50~$\Omega$. At room temperature these standards had less than 25~dB reflections across the frequency band between 300~kHz and 18~GHz. In the specific measurement band of interest in Fig.~\ref{fig:noise_twpa}, all standards had lower than 35~dB room temperature return loss. We detect a $4\mbox{ }\%$ increase in the mean of the transmission line phase velocity of the calibration standards after they are cooled to cryogenic temperature. This corresponds to a $4\mbox{ }\%$ increase in the reference characteristic impedance of Fig.~\ref{fig:noise_twpa}c that we attribute in this case to the decrease of capacitance during thermal contraction of the dielectric. Ongoing work focuses on improving the sensitivity of the cryogenic reflection measurement and establishing direct traceability of the cryogenic S-parameter measurements to metrological standards.}
%The minimum detectable reflection in the cryogenic measurement of Fig.~\ref{fig:source_schema}c is approximately -20 dB %due to external cabling in the fridge. Therefore we consider the S$_{11}$, and S$_{22}$ traces to represent upper bounds %of reflection from the attenuator within the noise source. 

\section{Traveling wave parametric amplifier}\label{appc}
\subsection{\label{appc1}Design and fabrication}
\begin{figure}[t]
\includegraphics{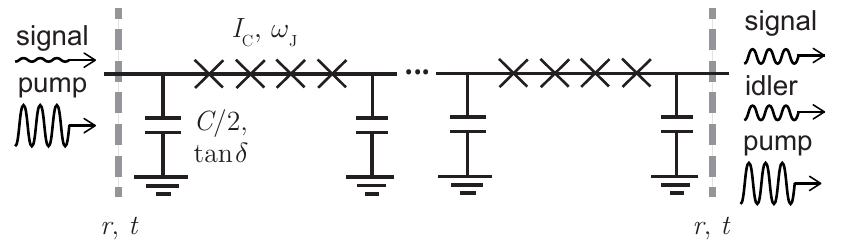}% Here is how to import EPS art
\caption{\label{fig:twpa_circ} TWPA electrical circuit. For the definition of symbols, see Table~\ref{tab:twpa_table}.}
\end{figure}

VTT has developed Josephson traveling wave parametric amplifiers (TWPAs) that are fabricated with a side-wall passivated spacer (SWAPS) niobium junction technology \cite{gronberg_side-wall_2017}. The multi-layer fabrication process relies on UV photolithography  and semi-automated 150-mm wafer processing steps. The substrate is high-resistivity silicon, and the layer thicknesses of the metals and the insulator range between 50~nm and 120~nm. As in Ref.~\citenum{white_traveling_2015}, the TWPA concept is a coplanar waveguide (CPW) transmission line where the center conductor is composed of Josephson junctions (Fig.~\ref{fig:twpa_circ}). The non-linearity of the Josephson inductance with respect to electric current allows a four-wave mixing process where the energy of a strong microwave pump tone (at frequency $f_\mathrm{p}$) is transferred to a weak signal (at frequency $f_\mathrm{s}$). At the same time, an idler tone emerges at the frequency $2f_\mathrm{p} - f_\mathrm{s}$. The non-linearity also causes a phase mismatch between the pump, signal, and idler, which could be compensated for by dispersion engineering \cite{white_traveling_2015,macklin_nearquantum-limited_2015}. In a lossless and phase-matched (-mismatched) TWPA the power gain grows exponentially (quadratically) with the device length.

\begin{figure}
\includegraphics{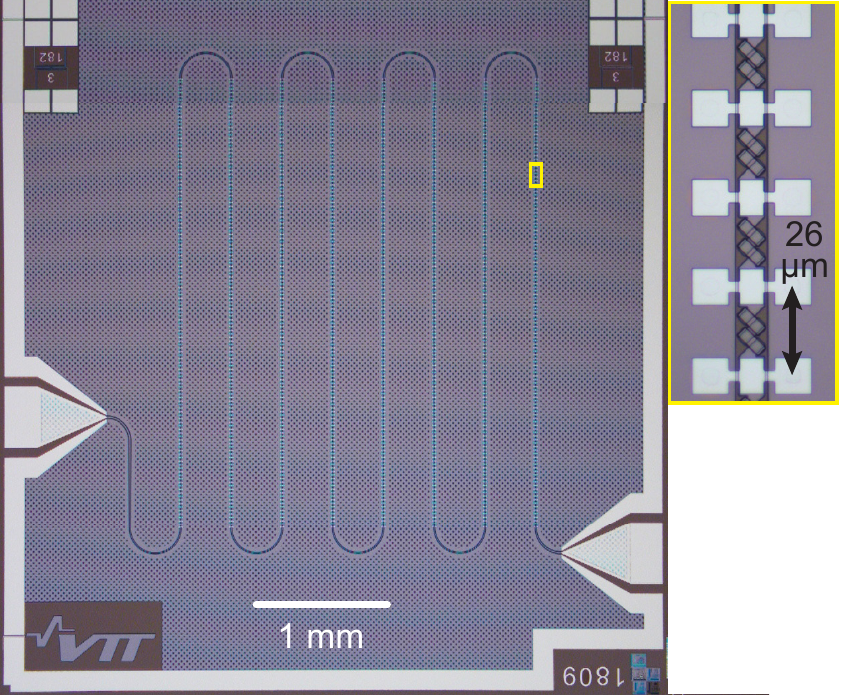}% Here is how to import EPS art
\caption{\label{fig:twpa_pic} Optical microphotographs of a TWPA chip which is nominally identical to the one that has been measured. The zoom-in shows about four unit cells. The light gray superconducting layer serves as the top electrode of parallel plate capacitors. Insulator contact holes (not clearly visible) on top of the CPW ground planes ensure that the top structure stitches together the ground potentials.}
\end{figure}

Microphotographs of our phase-mismatched TWPA show a meandering layout where a four-junction unit cell is repeated 1016 times (Fig.~\ref{fig:twpa_pic}). We target a critical current density of $100~\mathrm{A}/\mathrm{cm}^2$ for the Nb/Al-AlO$_x$/Nb junctions. The shunt capacitance to ground is dominated by parallel plates separated by atomic layer deposited AlO$_x$, with a capacitance density per unit area of about 1.5~fF$/\mu\mathrm{m}^2$. The electrical device parameters are listed in Table~\ref{tab:twpa_table}.

\ADDTXT{The TWPA chip with dimensions of 5~mm $\times$ 10~mm $\times$ 675~$\mu$m is mounted into a superconducting aluminum package. Inside the package there is a printed circuit board (PCB, material Rogers RO4003C) sandwiched between the aluminum base and lid. Aluminum bondwires connect the TWPA into the PCB. On the PCB, surface-mount subminiature push-on (SMP) microwave connectors provide an interface with coaxial wiring external to the package. Segments of ground-backed coplanar waveguide carry microwaves on the PCB between the SMP connectors and the TWPA chip.}

\begin{table}
\caption{\label{tab:twpa_table} TWPA device parameters.}
\begin{ruledtabular}
\begin{tabular}{ccc}
Parameter & Symbol & Value\\
\hline
Junction critical current& $I_\mathrm{c}$ & 4.4~$\mu{\mathrm{A}}$\\
Josephson plasma frequency~\footnotemark[1]& $\omega_J$ & $2\pi \times 46.5$~GHz\\
Unit cell physical length && 26~$\mu$m\\
Unit cell inductance (unpumped)~\footnotemark[2]& $L$ & 312~pH\\
Unit cell capacitance~\footnotemark[2]& $C$ & 115~fF\\ %
Dielectric loss tangent~\footnotemark[3]& $\tan \delta$ & 0.0025\\
Pump current amplitude~\footnotemark[4] & $I_\mathrm{p}$ & $0.53I_\mathrm{c}$\\
\hline
Mirror reflection coefficient~\footnotemark[4] & $r$ & 0.15\\
Mirror transmission coefficient & $t$ & $\sqrt{1-r^2}$\\
\end{tabular}
\end{ruledtabular}
\footnotetext[1]{The dispersiveness of the CPW is mainly determined by the Josephson plasma resonance and the unit cell $LC$ cut-off.}
\footnotetext[2]{Stray geometric inductance (capacitance) extracted from ANSYS HFSS electromagnetic simulations amounts to about 3~\% (4~\%) of the unit cell.}
\footnotetext[3]{In our numerical model only the signal and idler experience dielectric loss, while the pump remains unaffected.}
\footnotetext[4]{A free parameter when we optimize the numerical model.}
\end{table}

\subsection{\label{appc2}Numerical modeling}
To simulate the TWPA transmission we combine the following set of tools: coupled-mode equations (CMEs) of the co-propagating waves \cite{obrien_resonant_2014}, modeling of dielectric loss \cite{macklin_nearquantum-limited_2015}, and analysis of gain ripples \cite{planat_photonic-crystal_2020}. The pump is assumed to be stiff and the pump current has a constant amplitude $I_\mathrm{p}$. The magnitude of the idler entering the TWPA input is zero. The numerically solved CMEs produce the complex-valued transmission scattering parameters $\Son$ and $\Soff$ of the signal tone in a pumped and unpumped CPW, respectively. Imperfect radio-frequency mode conversions at the interfaces between the CPW and the sample holder are the principal reason for the gain ripples that we observe experimentally. We model the impedance mismatch with Fabry-Perot cavity formalism where mirrors (reflection coefficient $r$, transmission coefficient $t$) reside at the TWPA input and output (Fig.~\ref{fig:twpa_circ}) \cite{planat_photonic-crystal_2020}. In Fig.~3 we present simulated curves of the transmitted power along with the experimental data. They have been calculated as the squared absolute values of
\begin{subequations}
\begin{eqnarray}
\frac{t^2 \Son}{1-r^2 \Soff \Son}, \label{eq:appa3a}
\\
\frac{t^2 \Soff}{1-r^2 \Soff^2}. \label{eq:appa3b}
\end{eqnarray}
\end{subequations}
Here, Eq.~(\ref{eq:appa3a}) [Eq.~(\ref{eq:appa3b})] represents the pumped (unpumped) TWPA. Backward-propagating signals are assumed to experience $\Soff$.

In the numerical model we neglect the small geometric inductances and capacitances, the magnitudes of which are presented in Table~\ref{tab:twpa_table}. Since they are linear, low-loss circuit elements, the values of $I_\mathrm{p}$  and $\tan \delta$ are slightly underestimated.

\section*{Data availability}
The data that support the findings of this study are available from the corresponding author upon reasonable request.

% \begin{table}
% \end{tabular}
% \end{table}


%\pagebreak

%\bibliography{noise_references}

\end{document}